# New ion energy-mass spectrometer ULTIMAN (UlTImate Mass ANalyzer) for space plasmas

O.Vaisberg and S.Shuvalov

## Abstract

Measurements of Ion velocity distributions are one of basic goals of space plasma studies. There is variety of ion and electron spectrometers (e.g. Wuest et al, 2007, Young et al., 2007, Zurbuchen and Gershman, 2016, Vaisberg et al, 2016). The most frequently used ion spectrometer is top-hat analyzer (Carlson et al., 1983) consisting of toroidal electrostatic analyzer, electrostatic scanner and time-of-flight section with thin foil as start element and ion pre-acceleration.

We describe new energy-mass analyzer with electrostatic scanner providing hemispheric field-of view with small aberration, toroidal electrostatic analyzer, time-of-flight synchronizer with simple gate. It provides desirable hemisphere scanning, wide energy range and reasonable mass resolution. It can provide detailed measurements of the ion velocity distribution of ion species without significant gaps to obtain reliable moments of plasma flow. This analyzer without TOF ability can be used for measurements of electron component. With simple electro-optics elements this analyzer can be easily modified to many goals of plasma investigations.

## Introduction

Measurements of the space plasma are important element of in-situ investigation of the solar system including the solar wind, its interaction with planets and other bodies of the solar system, planetary magnetospheres and its losses, etc. The most often used instrument for measurements of hot plasmas in space is top-hat (Carlson et al., 1983) with nearly full coverage of the sphere and modest mass resolution. One of this instrument advantages is its many years heritage that encourage space agencies to give support to teams using this instrument.

Requirements of some space projects force ones to develop specialized plasma analyzer. Some examples are: (1) ion spectrometer PLASTIC for project STEREO that is measuring ion and ionization spectra of the solar wind ions with three techniques: electrostatic analysis, time-of-flight analysis and total energy measurements (Galvin et al., 2008); (2) the set of 8 top-hat analyzers on rapidly rotating spacecraft for project MMS to measure ion and electron velocity distribution with extremely high temporal resolution (Pollock, 2015); (3) energy-mass-analyzer with 2π field-of-view providing instant hemispheric measurement of ion flux at giving energy and energy scan (Vaisberg et al., 2016).

## Description

The scheme of new energy-mass analyzer if shown in Figure 1. It is cross-section of cylindrical SIMION (Manura and Dahl, 2008) model including multi-electrode angular scanner 1, entrance window 2 with gate electrode 3, toroidal electrostatic analyzer 4, micro-channel-plate detector 5, and ion trajectories 6. Trajectories shown are for viewing cone of 45° when potentials at scanner electrodes are 0. The beam shown consists of ions covering full velocity space of analyzer. Parameters of the model are fitting the MCP detector (shown as available from BASPIC enterprise) with sensitive area diameter 27 mm.

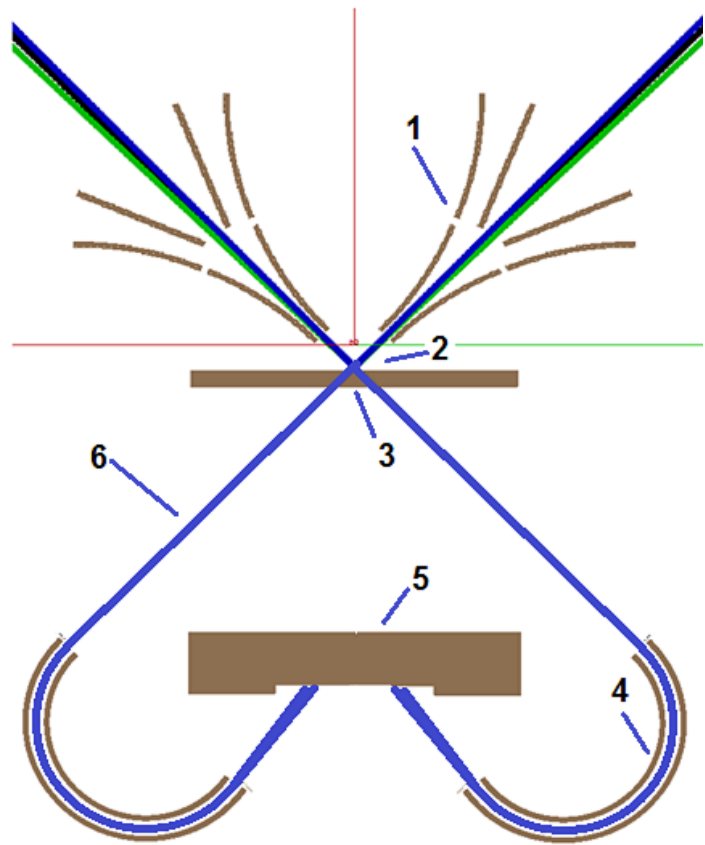

Figure 1. SIMION model of analyzer. 1-angular scanner, 2 – enters window, 3 – gate electrode, 4 – toroidal analyzer, 5 – MCP detector, 6 – ion trajectories.

The entrance of analyzer is the conic hole 2 with the cone 3. This cone is a part of entrance widow defining ion beam that enters the electrostatic analyzer. In the mass analysis mode the cone is at potential that deflects the ions from entering toroidal ESA. Sh ort grounding of potential at the cone generates short ion beam for time-of-flight mode of analyzer. The ion beam 6 enters the toroidal analyzer 4 which selects the ions with energy per charge, E/Q, determined by analyzer geometry and potentials applied to its electrodes. Selected ions are counted by MCP detector 5.

The scanning system 1 is selecting the cone angle of ions selected by analyzer. Variation of voltages applied to electrodes of scanner allows analysis of the ion flux within hemisphere (figure 2).

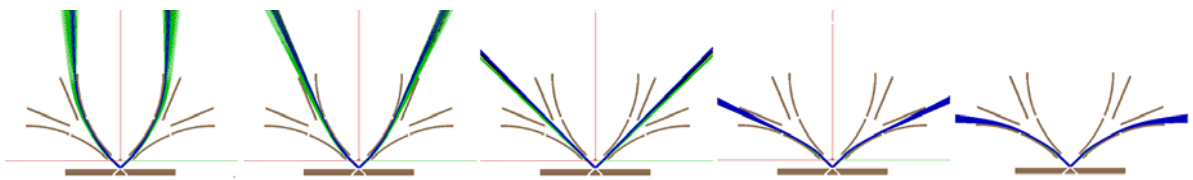

Figure 2. Work of scanner system within hemisphere with angular intervals 36° with appropriate potentials on scanner elements (all potentials are zeros for 45° cone).

## Performance simulation

In order to determine physical parameters of the energy-mass spectrometer a computer simulation was performed with ions launched from the surface located at some distance upstream of the instrument's aperture shown in Figure 3. This surface was a truncated cone (figure 3). Energy and initial velocity direction of generated particles had random uniform distributions over ranges $\Delta E = E_{max} - E_{min}$,

$\Delta\theta = \theta_{max} - \theta_{min}$, $\Delta\varphi = \varphi_{max} - \varphi_{min}$ where $E$ is energy, $\theta$ is polar angle and $\phi$ is azimuth. The values $E_{max}$, $E_{min}$, $\theta_{max}$, $\theta_{min}$, $\varphi_{max}$, $\varphi_{min}$ were set to cover the entire energy and angular band-pass of the instrument. Note that the surface over which the particles are generated, as well as the $\theta_{max}$ and $\theta_{min}$ values, are unique for each deflection state of electrostatic scanner.

The simulation was performed for the 45° deflection state of electrostatic scanner, when zero potentials are applied to the deflection electrodes. The parameters of simulation were: $E_{min} = 980\ eV$, $E_{max} = 1110\ eV$, $\theta_{min} = 43°$, $\theta_{max} = 47°$, $\Delta\varphi = 6°$. The number of generated particles $N_{in} = 2 \times 10^6$ was set to be sufficient enough for the error of the simulation to be less than 0.05% (the explanations are giver further in text).

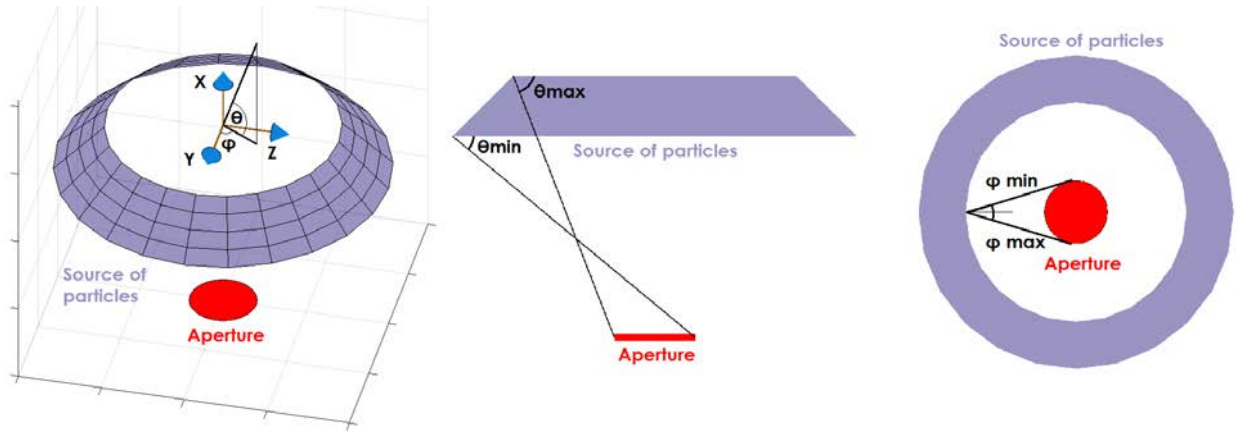

Figure 3. Initial conditions for generated particles. Dimensions are not to scale.

## Energy resolution

The energy distribution of $C = 54459$ ions which reached the detector is shown in Figure 4. The full width of distribution at half maximum is $\Delta E = 62\ eV$ which for the maximum location of the peak at $E = 1038\ eV$ gives the energy resolution very close to 6 %.

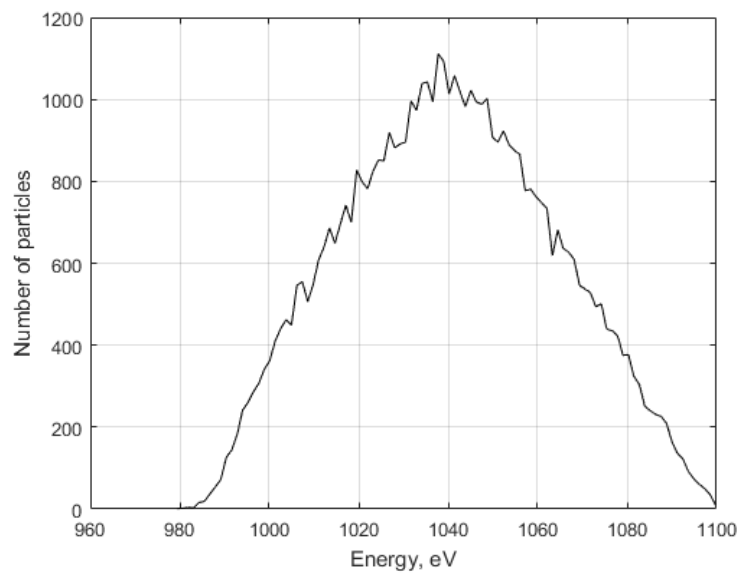

Figure 4. Curve of the energy band-pass of the instrument.

## Geometric factor

The equation for geometric factor for single energy and angular bin (omitting non-geometrical effects such as detection efficiency and transparency of grids) is as follows:

$$GF = S\Omega \cdot \frac{C}{N_{in}} \cdot \frac{\Delta E}{E}$$

Where $GF$ is geometric factor of the instrument, $S$ is the surface area on which particles are generated, and $\Omega$ is the solid angle observed from the instrument's aperture, and the ratio $\frac{\Delta E}{E}$ is the specific ratio allowing for band-pass of analyzer.

For polar angle $\theta = 45°$, the surface selected for particle generation is $S = 83 cm^2$. This yields in $S\Omega = S \int_{0°}^{6°} \int_{43°}^{47°} \sin\theta d\theta\, d\varphi \approx 0{,}429\ cm^2 ster$. As in our simulation $\frac{C}{N_{in}} = 2.72 \times 10^{-2}$ and $\frac{\Delta E}{E} = 6\%$ for the analyzer, we arrive at $GF = 6.73 \times 10^{-4}\ \frac{cm^2 \cdot ster \cdot eV}{eV}$.

## Mass resolution

In order to estimate the mass resolution of the instrument from computer simulation, a beam of particles with 40 u mass was launched from gate towards the detector. Initial energy and direction of particles were uniformly distributed over ranges wide enough to cover the whole band-pass of the instrument. A time of flight (TOF) **t** of each particle that reached the detector was recorded at the moment of its registration by model detector. Then, a histogram of recorded particles' time of flight was plotted (see figure 5), where time scale was converted to the scale of mass according to the rule $m = kt^2$, where m is the mass of particle, and k is the experimentally calculated coefficient. The resulting mass resolution is $\frac{m}{\Delta m} \approx 57$, where $\Delta m$ is full width at half maximum of the peak in figure 5.

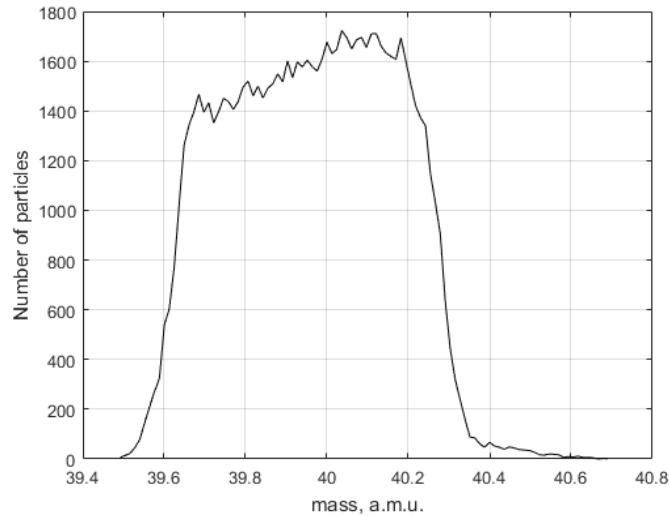

Figure 5. A histogram of TOF of particles with 40 a.m.u. mass which reached the detector. Time scale is converted to the scale of mass.

## Conclusion

Described energy-mass analyzer can be the working horse for the space missions intended to study the solar wind interaction with solar system bodies and planetary plasmas. It can provide full 3D velocity space distributions of ions with adequate energy range (~3 eV-30 keV) and mass resolution sufficient for analysis of main and secondary ion components. Expected mass of the instrument with the small number of electro-optics elements is relatively low (within 2 kg range for one hemispheric FOV analyzer and ~ 3 kg with 2 hemispheric analyzers with one electronics unit). Electro-optics of spectrometer allows expanding field of view in excess of 3π giving sufficient view for many applications. Initial model that is described in this manuscript suggests further refinement. Described electro-optics can be also used for electron spectrometer with smaller mass, as it does not require mass-resolution capability and due to higher electron flux in most circumstances. The results of laboratory model tests will be presented shortly for publication in scientific literature.

## References

Carlson, C.W., D.W. Curtis, G.Paschmann, and W.Michael, An instrument for rapidly measuring plasma distribution functions with high resolution, Adv. Space Res., Vol. 2, No. 7, 67-70, 1983.

Galvin, A. B., L. M. Kistler, M. A. Popecki, C. J. Farrugia, et al., The Plasma and Suprathermal Ion Composition (PLASTIC) Investigation on the STEREO Observatories, Space Science Reviews, April 2008, Volume 136, Issue 1–4, pp 437–486

Manura, D.J. and Dahl, D.A., *SIMION. Scientific Instrument Services*, Inc. Idaho National Laboratory, Revision, 2008.

Pollock, C., T.Moore, A.Jacques, J.Burch, et al., Fast plasma investigation for Magnetospheric Multiscale, Space Sci. Rev., 2016, DOI 10.1007/s11214-016-0245-4

Vaisberg, O., J.-J.Berthellier, L.Avanov, A.Leibov, F.Leblanc, F. Leblanc, P.Moiseev, D.Moiseenko, J. Becker, T Moore, M.Collier, J.Burch, D.McComas, J.Keller, G.Koynash, G.Lacky, H. Lichtenneger, R.Zhuravlev, A.Shestakov, S.Shuvalov, Panoramic charged particles analyzer based on 2π electrostatic mirror: all-sky camera concept and development for space missions. J. Geophys. Res. Space Physics, 121, doi:10.1002/2016JA022568.

Young, D.T., J.E.Nordholt, J.L.Burch, D.J.MsComas, et al., Plasma experiment for planetary exploration (PEPE), Space Sci. Rev., doi: 10.1007/s11214-007-9177-3, 2007.

Wuest, M.D.S.Evans, J.P.McFadden, W.I.Kasprzak, L.H.Brace, B.K.Dichter, W.R.Hoegy, A.J.Lazarus, A.Masson, and O.Vaisberg, Review of Instruments, in: Calibration of Particle Instruments in Space Physics, ed. By M.Wuest, D.S.Evans, and von Stieger, ISSI Scientific Report SR-007, pp.11-116, 2007.

Zurbuchen, T.H., and D.J.Gershman, Innovations in plasma sensors, J.Geophys. Res. Space Physics, 121, doi:10.1002/2016JA022493.